\documentstyle [12pt,epsf] {article} 
 
\catcode`@=12 
\begin{document} 
\vspace{0.5in} 
\oddsidemargin -.375in 
\newcount\sectionnumber 
\sectionnumber=0 
\def\be{\begin{equation}} 
\def\ee{\end{equation}} 
\thispagestyle{empty} 
\begin{flushright} 
UdeM-GPP-TH-02-100 \\ 
July 2002\\ 
\end{flushright} 
\vspace {.5in} 
\begin{center} 
{\Large \bf{Single Top Production 
\\ }} 
{\Large \bf{and Extra Dimensions\\}} 

\vspace{.5in} 
{ \rm Alakabha Datta{\footnote{email: datta@lps.umontreal.ca}\\}} 
{\it Lab Rene. J. -A. Levesque, 
Universite de Montreal,\\ 
C.P 6128 succrsale centre-ville\\
QC, H3C 3J7, Canada\\} 
\vskip .5in 
\end{center} 
\begin{abstract}
In extra dimension theories, where 
the gauge bosons of the standard model propagate in the bulk of 
the extra dimensions, there are 
 Kaluza-Klein excitations of the standard model gauge bosons that can 
couple to the standard model fermions. In this paper we study the effects 
of the excited Kaluza-Klein modes of the $W$ on single top production 
at the Tevatron.
\end{abstract}
\baselineskip 24pt  
Single top production is a very useful probe of the electroweak properties 
of the top quark which is believed to be intimately connected to 
electroweak symmetry breaking. For instance single top production offers 
the possibility of
measuring  the Cabibbo-Kobayashi-Maskawa (CKM) matrix element $V_{tb}$,
which is constrained by the unitarity of the CKM matrix  to be
 $ \sim 1$, and the the polarization of the top quark can
probe the $V-A$ nature of the weak interaction
\cite{Mahlon}. 
Single top production  
occurs within the SM in three 
different channels, the $s$-channel $W^*$ production, 
$q \bar q' \to W^* \to t \bar{b}$   
the 
$t$-channel $W$-exchange mode, $b q \to t q'$  
(sometimes referred to as $W$-gluon 
fusion), and through $t W^{-}$ production. 
The theory of single top production in 
the Standard Model(SM) has 
been extensively studied in the 
literature \cite{sm} while on the experimental front the
 D0 \cite{D0} and CDF \cite{CDF}
collaborations have already set limits on both the $s$-channel and $t$-channel
cross sections using data collected during run I of the Fermilab
Tevatron, and one expects 
discovery of these modes at the current
run II.
Measurement of the production cross section and
of single top quarks is also planned at  the CERN Large Hadron Collider (LHC)
\cite{Beneke:2000hk}.  
 
The large mass of the top quark, comparable to the electroweak symmetry
breaking scale, which makes the top quark sector very sensitive to 
new physic \cite{Peccei}. effects of new physics in singletop production 
have been studied extensively\cite{newphysics}. 
In this talk based on \cite{Datta:2000gm} 
 we consider 
effects that extra dimension theories can produce in single top production 
at the Tevatron. 
If in such theories, the gauge fields of the 
Standard Model(SM) live in the bulk of the extra dimensions 
then they will have Kaluza-Klein(KK) excitations  and the KK excited $W$ 
will contribute to single top production. 
To study the physics of the KK excited $W$ we use a model  
which is based on a simple extension of the 
SM to 5 dimensions (5D) \cite{edim}. However, as discussed above, we
 do not assume that this model represents all the physics 
beyond the standard model. The 5D SM is probably a part of a more
fundamental underlying theory.
The details of this model can be found in Ref.\cite{edim}

\begin{figure}[htb] 
\centerline{\epsfysize 6.2 truein \epsfbox{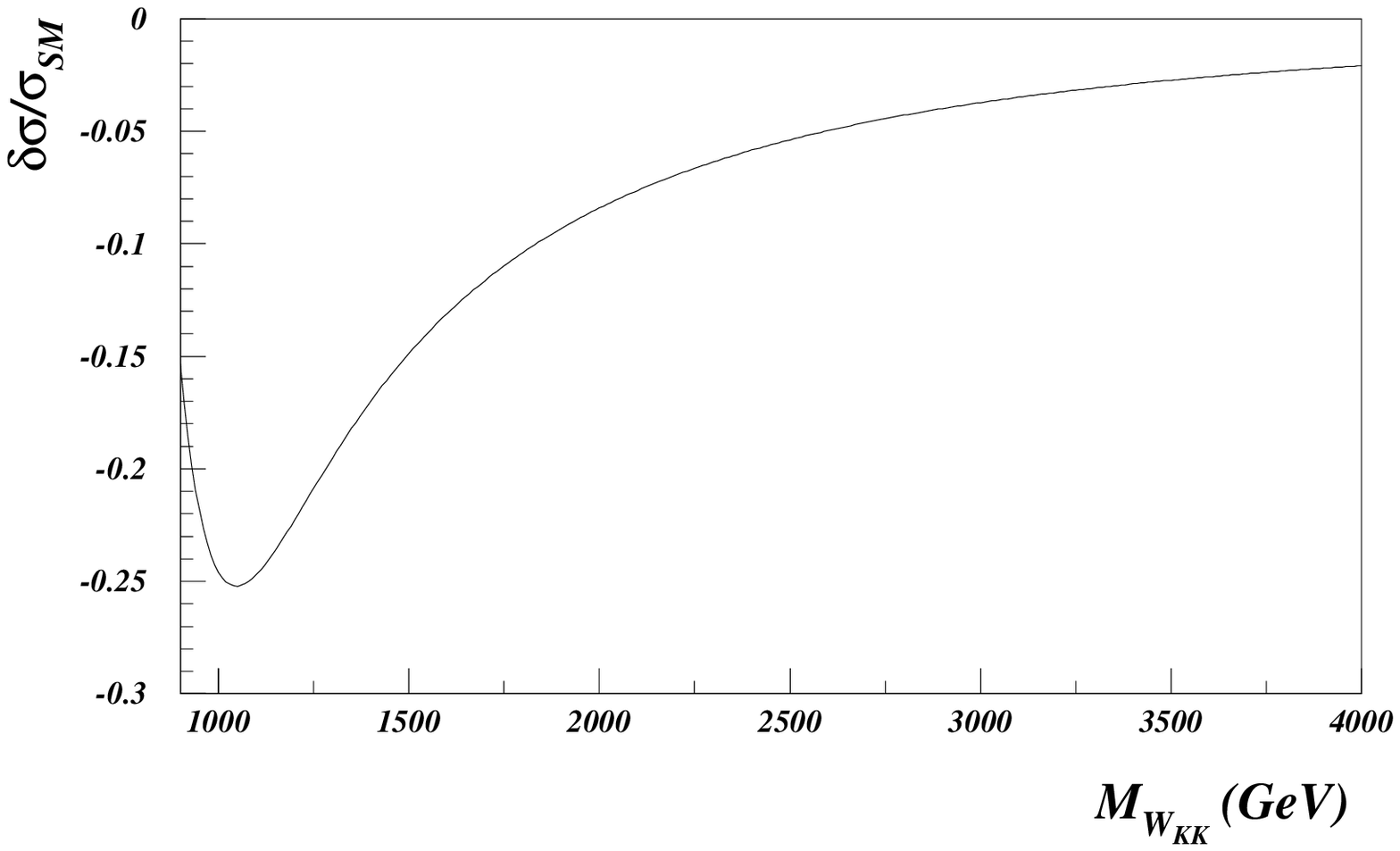}} 
\caption{$\Delta \sigma/\sigma_{SM}$ versus $M_{W_{KK}}$ the mass of the first KK excited 
$W$. } 
\end{figure} 
The  
piece of the effective Lagrangian, obtained after integrating over 
the fifth dimension, that is  relevant to our calculation  
is given by 
\begin{equation} 
{\cal{L}}^{ch}=\sum_{a=1}^2 {\cal{L}}_a^{ch} +{\cal{L}}_{new}
\end{equation} 
with 
\begin{eqnarray} 
{\cal{L}}_{a}^{ch}&=& 
\frac{1}{2}m^{2}_W 
W_a\cdot W_a 
+\frac{1}{2}M_c^2\sum_{n=1}^\infty \, n^2 \, W_a^{(n)}\cdot W_a^{(n)} 
\nonumber\\ 
&-&g\,W_a\cdot J_a-g\,\sqrt{2} J^{KK}_a\cdot \sum_{n=1}^\infty W_a^{(n)}\, , 
\end{eqnarray} 
where $m^{2}_W=g^2v^2/2$, the weak angle $\theta$ is defined by 
$e=g\, s_\theta=g'\, c_\theta$, while the currents are 
\begin{eqnarray} 
\label{currents} 
J_{a\mu}&=&\sum_\psi \bar{\psi}_L \gamma_\mu \frac{\sigma_a}{2}\psi_L\, , 
\nonumber\\ 
J_{a\mu}^{KK}&=& \sum_\psi \varepsilon^{\psi_L}\bar{\psi}_L \gamma_\mu 
\frac{\sigma_a}{2}\psi_L\, . 
\end{eqnarray} 
Here $\varepsilon^{\psi_L}$ takes the value 
1(0) for the $\psi_L$ living in 
the boundary(bulk). The mass of the $n^{th}$ excited KK state of the 
$W$ is given by $nM_c=n/R$ where R is the compactification radius. In this work we 
consider only the $n=1$ state. The term $ {\cal{L}}_{new}$ represents the additional new
physics beyond the 5 dimensional standard model the structure of which remains
unknown till the full underlying theory is understood. The coupling of
KK excited $W$ to the standard model is determined in terms of the
Fermi coupling, $G_F$, up to corrections of $O(m_Z^2/M_c^2)$ \cite{edim} . For $M_c \sim $ TeV
the $O(m_Z^2/M_c^2)$ effects
 are small for single top production and therefore we do not include
these effects in our calculations. We have  ignored the mixing of the
$W$ with $W_{KK}$ which is also an $O(m_Z^2/M_c^2)$ effect. Thus, assuming
 the $W_{KK}$
decays only to standard model particles, the predicted effect of $W_{KK}$
on single top production depends, in addition to the SM parameters, only on 
the unknown
mass of the $W_{KK}$.

The cross section for $p {\overline p}\rightarrow t{\overline b} X$ 
is given by 
\begin{eqnarray} 
\sigma(p {\overline p}\rightarrow t{\overline b} X) & = & 
\int dx_1dx_2[u(x_1){\overline d}(x_2)+ u(x_2){\overline d}(x_1)] 
\sigma(u{\overline d}\rightarrow t{\overline b}) . \ 
\end{eqnarray} 
Here $u(x_i)$, ${\overline d}(x_i)$ are the $ u $ and the ${\overline 
d}$ structure functions, 
$x_1$ and $x_2$ are the parton momentum fractions and the indices 
$i=1$ and $i=2$ refer to the proton and the antiproton. 
The cross section for the process 
$$u(p_1) + {\overline d}(p_2) \rightarrow W^* \rightarrow {\overline 
b}(p_3) + t(p_4) , $$ 
is given by 
\begin{eqnarray} 
\sigma & = & \sigma_{SM}\left[1 + 4 \frac{A}{D} +4 \frac{C}{D} \right], \nonumber\\ 
A & =& (s-M_W^2)(s-M_{W_{KK}}^2) + M_WM_{W_{KK}}\Gamma_{W}\Gamma_{W_{KK}}, \nonumber\\ 
C & = & (s-M_W^2)^2 +(M_W\Gamma_{W})^2, \nonumber\\ 
D & = & (s-M_{W_{KK}}^2)^2 +(M_{W_{KK}}\Gamma_{W_{KK}})^2, \ 
\end{eqnarray} 
and 
\begin{eqnarray} 
\sigma_{SM} & = & \frac{g^4}{384 \pi} \frac{(2s+M_t^2)(s-M_t^2)^2}{s^2[ 
(s-M_W^2)^2 +(M_W\Gamma_{W})^2]}. \ 
\end{eqnarray} 
Here $s=x_1x_2S$ is the parton center of mass energy while 
$S$ is the $ p{\bar p}$ center of mass energy . To calculate the width 
of the $W_{KK}$ we will assume that it 
decays only to the standard model particles. The $W_{KK}$ will then have the same 
decays as the $W$ boson but in addition it can 
also decay to a top-bottom pair which is kinematically forbidden for the $W$ boson. 
The width of the $W_{KK}$, $\Gamma_{W_{KK}}$, is then given by 
\begin{eqnarray} 
\Gamma_{W_{KK}} & \approx & 
\frac{2M_{W_{KK}}}{M_{W}}\Gamma_W+ 
\frac{2M_{W_{KK}}}{3M_{W}}\Gamma_W\cdot X ,\nonumber\\ 
X & = & ( 1-\frac{M_t^2}{M_{W_{KK}}^2}) 
( 1-\frac{M_t^2}{2M_{W_{KK}}^2} - \frac{M_t^4}{M_{W_{KK}}^4}). \ 
\end{eqnarray} 
where $\Gamma_W$ is the width of the $W$ boson and we have neglected the 
mass of the $b$ quark along with the masses of the lighter quarks and the 
leptons. 

In Fig. 1, we plot $\Delta \sigma/\sigma$ versus $M_{W_{KK}}$ , the mass of 
the first excited KK $W$ state, where $\Delta \sigma$ 
is the change in the 
single top production cross section in the presence of $W_{KK}$ 
and $\sigma$ is the standard 
model cross section\footnote{ We have 
not included the QCD and Yukawa corrections to the single top quark 
production rate. They will enhance the total rate, but not change 
the 
percentage of the correction of new physics to the cross section. 
}. 
We have used the CTEQ \cite{CTEQ} structure 
functions for our calculations and obtain a standard model cross section 
of 0.30 pb for the process $p {\overline p}\rightarrow t{\overline b} X$ 
at $\sqrt{S}= 2$TeV. 
We observe from Fig. 1 that the presence of $W_{KK}$ can 
lower the cross section by as much as 25 \% for $M_{W_{KK}} \sim 1 $TeV
The inclusion of the higher KK resonance can than lower the cross
 section by more than 40\%. 
This has an important implication for the measurement of $V_{tb}$ using 
the s-channel mode at the Tevatron. It was pointed out in Ref\cite{topflavor}
that there could be models where
 the presence of an additional $W$( denoted as
$W'$) could lead to a measurement of the 
cross section for the s-channel $p {\overline p}\rightarrow t{\overline b} X$ 
smaller than the standard model prediction. 
This could, as pointed out in 
Ref \cite{topflavor},
lead one to conclude that
$V_{tb} <1$ which could then be wrongly interpreted as 
evidence for the existence of new generation(s) of fermions mixed with the 
third generation. 
This work provides a specific example of such a model but
the key point is that if
$V_{tb}$ were less than unity, one
 would see a decrease of {\it{both}}  the s-channel and the t-channel 
cross section unlike the KK excited $W$ case whose effect would mainly be 
on the s-channel process and not so much on the
 the t-channel process. 
Furthermore, most realistic models with 
an extra $W'$ do not predict a large decrease in 
the s-channel cross section\cite{ES}.
Hence a large decrease of the s-channel cross section {\it{without}}
 a corresponding decrease in the t-channel cross section would indicate the presence of 
KK excited $W$ bosons.

{\bf Acknowledgment:} 
This work was supported in part by Natural Sciences and Engineering Research Council 
of Canada.

\end{document}